# RESONANT ATOM TRAPS FOR ELECTROMAGNETIC WAVES

V. Danilov

*Spallation Neutron Source Project, Oak Ridge National Laboratory, 8600 Bldg, MS 6462, Oak Ridge TN 37830*

**Abstract**

Exitation of atomic levels due to interaction with electromagnetic waves has been the subject of numerous works, both experimental and theoretical. This topic became of interest in accelerator physics in relation to high efficiency charge exchange injection into rings for high beam power applications. Taking equations of resonant atom-wave interaction equations as a basis, this paper shows that there exist some interesting phenomena which lead to the existence of trapped electomagnetic waves (photon traps) in a medium that consists of atoms with transition frequencies in proximity to the wave frequency. These traps may exist in random and periodic lattices, and may have very low loss rate. The atomic medium can serve as an excellent wavegiude or tool to form and transmit electromagnetic waves for applications to accelerators and to electromagnetic devices in general, where high pressure gas use is acceptable. In addition, such traps in gases may accumulate substantial energy for a long period of time, leading to the possibility of creating objects similar (or equivalent) to ball lightning.

## I.    INTRODUCTION

Periodic lattices of atoms, placed in a monochromatic electromagnetic wave with its frequency close to that of some transition frequency between atomic levels, produce either oscillatory solutions for the wave function in the space domain, corresponding to propagating waves, or exponentially growing/decaying solutions in the limit of zero electric field.  When the fields are strong, nonlinearity changes the nature of the interaction, giving rise to finite in space (or trapped) solutions for the coupled atom-wave equations. The equation for trapped waves in the Rotating Wave Approximation (RWA) (see, e.g., [1]) for a 1D lattice of simple hydrogen-like atoms is given in the next section. Section III introduces approximate solutions for these equations. Appendix A provides trapped wave numerical solutions for 1D random atom lattices. It is shown that the localization phenomenon, discovered by Anderson in connection with random lattices in metallic alloys (see [2]) may play a significant role in creation of trapped modes as well as in the nonlinearity of the atom-wave interaction. Section IV presents a generalization of the results for 2D and 3D cases for an averaged equation in the case of high atom density. Section V gives practical estimations for the achievable electric field of the trapped wave in the case of real media. The conclusion summarizes the results.

## II.    1D SEMICLASSICAL EQUATIONS FOR RESONANT ATOM-WAVE INTERACTION

To obtain the solution for trapped electromagnetic fields, the easiest approach is to use semi classical equations for two level atoms, interacting with a monochromatic classical electromagnetic wave.  The atoms may have more than two levels, but we consider here only the resonant case in which some two level transition frequency $\omega_0$ is close to that of the wave frequency $\omega$. In this section, we deal also with a periodic 1D lattice, consisting of straight sections, where the electromagnetic wave propagates freely, and of infinitely short 2D planes, consisting of atoms with area density $N$ arranged perpendicular to the wave propagation direction. The equation for slow varying[a] probability amplitudes $C_1$ (lower level) and $C_2$ (upper level) is (see [1]):

$$\dot{C}_1 = \frac{i\mu_{12}E_0}{2\hbar} C_2 e^{i\Delta t},$$
$$\dot{C}_2 = \frac{i\mu_{21}E_0}{2\hbar} C_1 e^{-i\Delta t}, \quad (1)$$

where $\Delta = \omega - \omega_0$, the electric field has the form $E = E_0 \cos \omega t$, $\mu_{12} = \mu_{21}^* = -\int d^3 r \, u_1^*(\vec{r}) ez u_2(\vec{r})$ (assuming the light is polarized in the direction $z$, parallel with the atomic plane), and $u_1$ and $u_2$ are the normalized wave functions of the lower and the upper excited states, respectively. In the case of hydrogen, the lower level has primary quantum number n=1. The upper level primary quantum number is determined by the resonant condition, its angular momentum l=1, and the projection of angular momentum on the z axis m=0. For this case $\mu = \mu_{1n}$ is real and we omit its subscripts for simplicity.

These equations have an interesting solution in which both amplitudes $C_1$ and $C_2$ oscillate at fixed frequencies, with no change of their absolute values:

$$C_{1,2} = |C_{1,2}| \exp(i\omega_{1,2} t). \quad (2)$$

Substituting (2) to (1) yields:

$$\omega_1 = \omega_2 + \Delta,$$
$$\omega_2 = \frac{-\Delta \pm \sqrt{\Delta^2 + \Omega^2}}{2}, \quad (3)$$

where $\Omega = \mu E_0 / \hbar$ is the Rabi frequency at exact resonance. It is clear that the atomic excitation is constant in time for these solutions and that the electric current, caused by oscillations of probability amplitudes, has a 90 degree phase shift with respect to the electric field. Thus, the energy of the field is neither absorbed nor emitted by the atoms. To calculate atomic reflection and transmission of the electromagnetic wave, we use formulas for the atomic plane polarization $P_z$, induced by the electric field, and Maxwell's equations for the electric field, influenced by atomic currents, from [3]:

---

[a] Slow varying amplitudes means that the real amplitudes are slow amplitudes, multiplied by $exp(-i\omega_i t)$, where $\omega_i$ is the corresponding level eigenfrequency.

$$P_z(x,t) = 2N\mu \operatorname{Re} C_{1f} C_{2f}^* \delta(x),$$

$$\left(\frac{\partial^2}{\partial x^2} - \frac{\partial^2}{c^2 \partial t^2}\right) E_z = \frac{1}{\varepsilon_0 c^2} \frac{\partial^2 P_z(x,t)}{\partial t^2}, \tag{4}$$

where $N$ is the area density of atoms, the atomic plane is located at $x=0$, and all atoms have same frequency. The subscript $f$ for probability amplitudes means we have to use their "fast" values and real frequencies of levels (see footnote on page 2). Now we are in position to calculate the transmission and reflection coefficients of the wave at the atomic plane. The electric field at $x=0$ is $E=E_0 \cos\omega t$, and the solution of (1) for "slow" probability amplitudes becomes:

$$C_1 = \pm \frac{1}{\sqrt{1 + \left(\frac{\Omega^2}{4\omega_2^2}\right)}} \exp(i\omega_1 t),$$

$$C_2 = \pm \frac{\Omega}{2\omega_2 \sqrt{1 + \left(\frac{\Omega^2}{4\omega_2^2}\right)}} \exp(i\omega_2 t). \tag{5}$$

Multiplying by factors $\exp(-i\omega_{l1,l2} t)$, their own Eigen frequencies, and substituting into (4), the polarization $P_z(x,t)$ becomes:

$$P_z(x,t) = \frac{N\mu\Omega}{\varepsilon_0 \omega_2 \left(1 + \left(\frac{\Omega^2}{4\omega_2^2}\right)\right)} \cos(\omega t) \delta(x) = \frac{N\mu^2 E(x=0)}{\varepsilon_0 \hbar \omega_2 \left(1 + \left(\frac{\Omega^2}{4\omega_2^2}\right)\right)} \delta(x), \tag{6}$$

where $\omega$ is the electromagnetic wave frequency. One can see that the polarization oscillates either in phase or anti phase with electric field, depending on sign of $\omega_2$. The current, which is proportional to the polarization derivative, oscillates with a phase shift of 90 degrees relative to the field phase. To calculate the transmission and reflection coefficients, we take the electric field coming from one (positive $x$) side of the atomic plane as a sum of incident $E_i$ and reflected $E_r$ waves. On the other side of the plane we have, for simplicity, only the transmitted wave $E_t$. In mathematical notation, they are:

$$\begin{aligned}
E_{i+} &= E_0 \cos k(x + ct), \\
E_{r+} &= E_1 \sin k(x - ct) + E_2 \cos k(x - ct), \\
E_{t-} &= E_3 \cos k(x + ct) + E_4 \sin k(x + ct),
\end{aligned} \tag{7}$$

where $E_i$ are the amplitudes of the waves, $k=\omega/c$, and the signs + and - stand for positive and negative $x$ coordinate. From equations (4) it follows that the electric field is continuous at *x=0*, and the derivative of the electric field with respect to x jumps $\dfrac{Nk^2\mu^2 E(x=0)}{\varepsilon_0 \hbar \omega_2 (1+(\dfrac{\Omega^2}{4\omega_2^2}))}$.

This yields:

$$E_0 + E_2 = E_3,\ E_1 = -E_4,$$

$$(E_1 - E_4)k = \frac{Nk^2\mu^2 E_3}{\varepsilon_0 \hbar \omega_2 (1+\dfrac{\Omega^2}{4\omega_2^2})}, \qquad (8)$$

$$(E_2 - E_0 + E_3)k = \frac{Nk^2\mu^2 E_4}{\varepsilon_0 \hbar \omega_2 (1+\dfrac{\Omega^2}{4\omega_2^2})},$$

where $\omega_2, \Omega$ should be taken at the electric field amplitude $\sqrt{E_3^2 + E_4^2}$ which is the modulus of the electric field at *x=0*.

Equations (7-8) can be rewritten in more compact complex form:

$$E_{f-} + E_{b-} = E_{f+} + E_{b+},$$

$$(E_{f+} - E_{f-}) = -\frac{iNk\mu^2 (E_{f-} + E_{b-})}{2\varepsilon_0 \hbar \omega_2 (1+\dfrac{\Omega^2}{4\omega_2^2})}, \qquad (9)$$

where the complex amplitudes of the forward and backward waves are related to real fields through $E_{real} = \mathrm{Re}\, E_{f(b)} \exp(-ik(ct \mp x))$ [b]. The subscripts *f* and *b* denote the forward and backward directions (forward means the direction to +infinity), and the subscripts + and – are related to the fields immediately to the right and to the left of the atomic plane, respectively. One can check that equations (9) are valid when we take the incident wave in (7), coming from the other direction. When one moves one step in the positive direction from one lattice node to another, the complex amplitudes of the forward and backward waves are multiplied by factors $\exp(-i\phi)$ and $\exp(i\phi)$, respectively, where $\phi$ is the wave phase advance between the two atomic planes. Combining these factors with (9), we obtain the transformation of the wave complex amplitudes over one lattice period:

---

[b] The sign of factor in front of *t* in the exponent should be the same for forward and backward waves to satisfy (8) and (9).

$$E_{f,n+1} = \exp(-i\phi)(E_{f,n} - \frac{iNk\mu^2(E_{f,n}+E_{b,n})}{2\varepsilon_0\hbar\omega_2(1+\frac{\Omega^2}{4\omega_2^2})}),$$

$$E_{b,n+1} = \exp(i\phi)(E_{b,n} + \frac{iNk\mu^2(E_{f,n}+E_{b,n})}{2\varepsilon_0\hbar\omega_2(1+\frac{\Omega^2}{4\omega_2^2})}),$$

(10)

where we must take $\omega_2, \Omega$ at the electric field amplitude $|E_{f-}+E_{b-}|$. The index $n$ stands for the lattice node with ascending numbers corresponding to the "right" direction. It is convenient to normalize the electric field such that its Rabi frequency is expressed in terms of frequency offset $\Delta$. The new variable for this normalization is $u = \frac{2\mu E}{\hbar\Delta}$. This transforms the equations (10) to the following form:

$$u_{f,n+1} = \exp(-i\phi)(u_{f,n} \mp \frac{i\kappa(u_{f,n}+u_{b,n})}{2\sqrt{1+|\frac{u_{f,n}+u_{b,n}}{2}|^2}},$$

$$u_{b,n+1} = \exp(i\phi)(u_{b,n} \pm \frac{i\kappa(u_{f,n}+u_{b,n})}{2\sqrt{1+|\frac{u_{f,n}+u_{b,n}}{2}|^2}}),$$

(11)

where $\kappa = \frac{Nk\mu^2}{\varepsilon_0\hbar\Delta}$.

To find stationary solutions, we can further simplify this map when we have $u_f$=c.c. $u_b$ in some straight section. One can check that the map (11) preserves this relation of the waves: the forward wave complex amplitude will always be the complex conjugate of the backward wave amplitude. Taking *Re* $u_f$=q, *Im* $u_f$=p we have:

$$q_{n+1} = q_n \cos\phi + (p_n \mp \frac{\kappa q_n}{\sqrt{1+|q_n|^2}})\sin\phi,$$

$$p_{n+1} = -q_n \sin\phi + (p_n \mp \frac{\kappa q_n}{\sqrt{1+|q_n|^2}})\cos\phi.$$

(12)

Map (12) is a typical 1D symplectic map of classical mechanics. The phase space of this motion has stable and unstable points, closed invariant curves, chaotic trajectories, etc. Here we concentrate mostly on the behavior of the fields near the phase space center point with coordinates (0,0).

For small $\kappa \sim 1$ the behavior of the solution near zero is of two types: stable motion around the center, and unstable motion, which approximately follows a closed curve away from and returning to the coordinate center (0,0). Figure 1 shows the map's (12) (with the sign chosen negative in the first and second equation) electric field behavior over the lattice nodes with initial conditions $(x,p)=(0.00002, 0.00002)$ and parameter $\kappa=-0.5$ and $\phi=1.0$ radian. When the parameter $\kappa$ increases, the center point (0,0) becomes unstable. Linearizing (12) around zero, and finding eigenvalues of the matrix, one gets the condition for stability:

$$|\cos\phi \mp \frac{\kappa}{2}\sin\phi| < 1. \qquad (13)$$

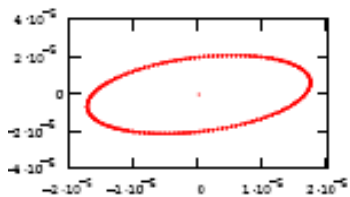

**FIG. 1.** Electric field oscillation in 1D lattice, consisting of resonant atom planes. Horizontal and vertical axes show real and imaginary part of the forward field, respectively. Parameter $\kappa = -0.5$.

Figure 2 shows a case when the center point is unstable. The parameters are $\kappa=-3.3$ and $\phi=1.0$ radian. The trajectory passes near a closed separatrix that emerges from and returns to the unstable fixed point in the center of the phase space. This description of the motion is not exact – in order to have a closed separatrix we need to have a "good" (e.g., analytic) integral of motion for the map (12). Such integrals of the motion, in general, do not exist. In our case we doubt that the motion is integrable. For large values of the parameter $|\kappa|>10$ the separatrix for the center point becomes fuzzy and finally the smooth loop-like trajectory breaks into a chaotic web around many resonant islands (see Figure 3 for a trajectory with the same initial conditions but with $\kappa=-53.3$).

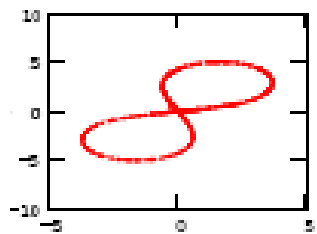

**FIG. 2.** Unstable behavior of electric field over resonant atomic lattice nodes starting from the vicinity of (0,0). Parameter $\kappa=-3.3$.

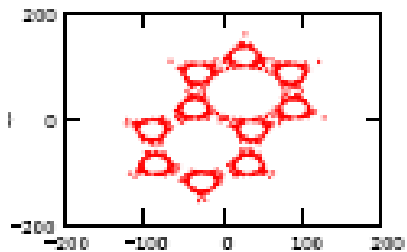

**FIG. 3.** Map (12) trajectory for large absolute values of $\kappa = -53.3$ starting from the vicinity of (0,0).

The separatrix in Figure 3 is in reality a very complicated heteroclinic structure, discovered by Poincare (see [4] for description of general separatrix motion for symplectic maps). In general, there exist outgoing and incoming separatrix curves for any fixed point. These curves cross each other with some angle; the crossing point travels away from fixed point and back to it in the case when crossing occurs for separatrices of the same resonance.

The points, traveling from and to the center point are very remarkable – they represent a trapped mode of the field that exponentially decays at "±" infinity. In the case of small $\kappa$ all the points of the separatrix return back or extremely close to the central fixed point. The next section will present an approximation of the map (12) by a differential equation, for which the motion turns out to be exactly integrable and the separatrix is a continuous line. Appendix A, however, presents a case with random lattice in which separatrix curves don't constitute a single line, but there exist crossing points that go from and come to the center point, representing a trapped mode with exponential decay of the electromagnetic field at both infinities.

### III.  APROXIMATE SOLUTION FOR 1D STATIONARY DISTRIBUTIONS

When $\phi$ and $\kappa$ are small, the map (12) can be rewritten in the following form:

$$q_{n+1} = q_n + p_n \phi,$$
$$p_{n+1} = -q_n \phi + p_n \mp \frac{\kappa q_n}{\sqrt{1+|q_n|^2}}. \qquad (14)$$

Keeping in mind that all changes of $p,q$ are small we can transform the finite difference equation (14) into a differential equation, approximating the map:

$$\frac{\partial q}{\partial x} = k\, p,$$
$$\frac{\partial p}{\partial x} = -k\, q \mp \frac{\kappa q}{l\sqrt{1+|q|^2}}, \qquad (15)$$

where $k=\omega/c$ and $l$ is the distance between lattice nodes. Now we pick the + sign in the "source" term and write down the Hamiltonian of the motion (note that coordinate $x$ here has meaning of time):

$$H = \frac{k(p^2+q^2)}{2} - \frac{\kappa}{l}\sqrt{1+q^2}, \quad (16)$$

Only conditions $\frac{\kappa}{l} > k$ lead to instability and trapped modes: for this condition the central point becomes hyperbolic. The equation for the separatrix reads:

$$p_{sep} = \sqrt{\frac{2\kappa}{kl}(\sqrt{1+q^2}-1) - q^2}, \quad (17)$$

where $q^2 < 1 - \frac{kl}{\kappa}$. Figure 4 shows the phase space of the Hamiltonian (16) for $\kappa = 2kl$. One can see two "eyes" with stable motion, a separatrix, corresponding to the trapped mode, and an outer region with oscillations around "two-eyed" region.

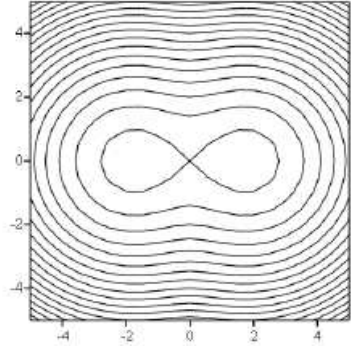

**FIG. 4.** Real (horizontal axis) and imaginary part (vertical axis) of the electric field in 1D lattice (arbitrary units). The separatrix corresponds to the trapped mode with exponential decay of the field at infinity.

The dependence of the normalized "brightness" (the square of normalized field at the center), and the trapped mode size on parameters that can easily be obtained from Hamiltonian picture. The normalized "brightness' is roughly the extent of the separatrix in the horizontal direction and is:

$$q^2 = 1 - \frac{kl}{\kappa}, \quad (18)$$

where $\frac{\kappa}{l} > k$. The real "brightness" is the normalized brightness multiplied by the parameter $(\frac{\hbar\Delta}{2\mu})^2$ (see the definition of "normalized field" right before Equation (11)). It is easy to see that the maximum electric field in our trapped modes corresponds to the case where the related Rabi frequency equals $\Delta$.

The "time" in our case is measured in units of the wavelength and the trapped state size is just the time for the field to travel from the maximum coordinate separatrix value to the point where the field is reduced by a factor of $e$. This is, roughly, a half period of oscillation around the stable point in the center of the "eye". By finding the stable fixed point, and linearizing the motion around it, one can easily find the half-period of the oscillations, which is the approximate size $S$ of the trapped state:

$$S = \frac{\lambda}{2}\sqrt{\frac{\kappa^2}{\kappa^2 - (kl)^2}}, \qquad (19)$$

where we have $\frac{\kappa}{l} > k$ again, and the $\lambda$ is the wavelength of the free electromagnetic wave. One can see that in the limit $\kappa \to kl$ the size approaches infinity but the "brightness" goes to zero.

## IV. TRAPPED MODES FOR 2D AND 3D CASES

Equation (4) can be generalized to two and three dimensions. Indeed, we can use the expression of the spatially averaged source term from (4) in Helmholtz equations. The density itself must be small enough to allow the neglect of terms involving the interaction of electron dipole moments (e.g., static dipole fields). Definitely, the model is not valid for solids – energy shifts of molecules due to electron interactions are too large to ignore. To satisfy this requirement, the density $N$ should be less than some parameter $K$ of the order of solid density. To derive the equation for 2D and 3D cases, we use here the equation for the Hertz vector $Z$ and the equation that relates it to the electric field (see, e.g., [6]):

$$\nabla^2 \vec{Z} + k^2 \vec{Z} = \pm \frac{1}{k^2} \frac{\chi \vec{E}}{\sqrt{1+|E|^2}},$$
$$\vec{E} = \nabla \operatorname{div} \vec{Z} - \frac{1}{c^2} \frac{\partial^2}{\partial t^2} \vec{Z}, \qquad (20)$$

where the new interaction parameter is $\chi = \frac{Nk^2\mu^2}{\varepsilon_0 \hbar \Delta}$, with $N$ now being a conventional atom density. These equations are complicated and rich in various solutions. We would like to reduce the number of cases for analysis by choosing only solutions with $div\ \vec{Z} = 0$ (and, consequently, $div\ \vec{E} = 0$). In this case the equations take the form of Equation (4), but for vector electric field:

$$\nabla^2 \vec{E} + k^2 \vec{E} = \pm \frac{\chi \vec{E}}{\sqrt{1 + |E|^2}}, \qquad (21)$$

$$div\ \vec{E} = 0.$$

One can see that the 1D lattice equations have the same form after substitution $\frac{\kappa k}{l} \to \chi$.

The first case we analyze has cylindrical symmetry with the electric field oriented along the $z$ axis. The equation for the field $E_z$ reads:

$$\frac{\partial^2 E_z}{\partial (kr)^2} + \frac{1}{kr} \frac{\partial E_z}{\partial kr} + E_z = \frac{\chi E_z}{k^2 \sqrt{1 + |E_z|^2}}, \qquad (22)$$

where we are interested in only positive $\chi$ and thus use the plus sign in front of the "source" term. Next, we only choose cases for which there is no propagation of waves to infinity $\chi > k^2$ [c], and we analyze only trapped modes, i.e. the fields exponentially decay at infinity. The other criteria for our choice of solutions is that they should be finite at $r=0$. Numerical analysis shows that there exist many (our guess is that there exist infinitely many) solutions for trapped modes. They can be enumerated by an integer index that corresponds to the number of each solution's zero crossings. Figure 5 shows functions $F_0$ (left) and $F_8$ (right) for the value $\chi / k^2 = 1.1$

---

[c] As was pointed out in the beginning of the section, the density shouldn't be too large for the model to be valid. At the same time, the gas should be dense enough to satisfy a necessary trapped mode condition $\chi > k^2$. Since the parameter depends inversely on difference of radiation and transition frequencies, the last condition can always be met.

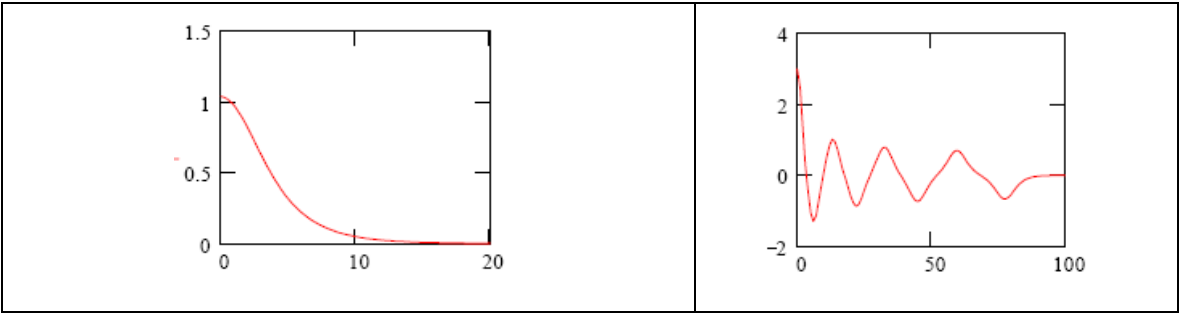

**FIG. 5.** $F_0$(left) and $F_8$(right) as a function of r.

One interesting feature of the solutions with large index is that the corresponding trapped mode size can be arbitrarily large. It was found numerically for solutions with low index that the size and "brightness" of the trapped states are similar to those for the 1D case (see equations (18) and (19)). Accurate numerical solutions can be easily obtained by simple computer programs and are omitted here. Instead, we will construct approximate numerical solutions for the trapped modes in the 3D case.

It is quite probable that the 2D cylindrical mode, if curved into a toroid, produces a 3D solution when the aspect ratio of the torus becomes large. To look for similar azimuthally symmetric solutions in spherical geometry, we use a particular form of equation (21) for $E_\phi$, independent of $\phi$:

$$\frac{\partial^2 E_\phi}{\partial r^2} + \frac{2}{r}\frac{\partial E_\phi}{\partial r} + \frac{1}{r^2 \sin\theta}\frac{\partial}{\partial \theta}\sin\theta \frac{\partial E_\phi}{\partial \theta} - \frac{E_\phi}{r^2 \sin^2\theta} = \frac{\chi E_\phi}{\sqrt{1+|E_\phi|^2}}, \qquad (23)$$

where $\theta$ is the polar angle. Unfortunately, there is no solution to this equation, independent of $\theta$. Therefore, we will look for a solution periodic in $\theta$, in the following general form:

$$E_\phi(r,\theta) = \sum_{n=1}^{\infty} f_n(r) P_n^1(\cos\theta), \qquad (24)$$

where $P_n^1(\cos\theta)$ are associated Legendre polynomials of index 1. The choice is obvious – the LHS of equation (23) is separable when we substitute Equation (24). Each term (with appropriate $f_n(r)$) becomes a solution to LHS of (23) since the polynomials obey the following equation:

$$\frac{1}{\sin\theta}\frac{\partial}{\partial \theta}\sin\theta \frac{\partial P_n^1}{\partial \theta} - \frac{P_n^1}{\sin^2\theta} = -n(n+1)P_n^1. \qquad (25)$$

The polynomials have the following orthogonality property:

$$\int_{-1}^{1} P_l^1(t) P_n^1(t) dt = \frac{2l(l+1)}{2l+1} \delta_{nl}, \qquad (26)$$

and the first few of them are: $P_1^1(\cos\theta) = -\sin\theta$, $P_2^1(\cos\theta) = -3\sin\theta\cos\theta$, $P_3^1(\cos\theta) = -\frac{3}{2}(5\cos^2\theta - 1)$, etc. Substituting (24) into (23), multiplying both sides by an associated Legendre polynomial with index $l$, and integrating the resulting equation using condition (26) over θ, yields:

$$\frac{\partial^2 f_l}{\partial r^2} + \frac{2}{r}\frac{\partial f_l}{\partial r} + f_l(k^2 - \frac{l(l+1)}{r^2}) = \frac{\chi(2l+1)}{2l(l+1)} \sum_{n=1}^{\infty} \int_0^{\pi} \frac{P_l^1(\cos\theta) P_n^1(\cos\theta) \sin\theta}{\sqrt{1 + (\sum_{n=1}^{\infty} f_n(r) P_n^1(\cos\theta))^2}} d\theta. \qquad (26)$$

As in the case of cylindrical solutions, we found many approximate solutions for some small harmonic numbers. One of these harmonic numerical solutions is shown in Figure 6. The figure shows the cross section of the trapped mode when y=0. One can see that it consists of two torroids (rings) one atop of another. Of course, finding numerical solutions is no proof that the procedure converges as the number of harmonics increases, but this approach works in most cases and we assume this is the case for our equations.

We guess that more spherical forms of solutions can be obtained when we try to find solutions for $E_z$ and $E_\theta$, rather than $E_\phi$. Unfortunately, in this case the equations become more involved and the analysis is left for future developments of the topic.

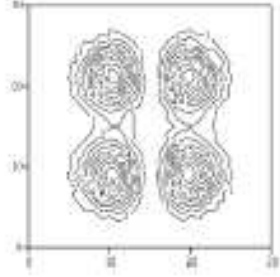

**FIG 6. Electric field squared in 3D trapped azimuthally symmetric mode as a function of X, Z coordinates. Shown is the cross section of the mode with Y coordinate equal to zero.**

In addition to the stationary distribution, one may look for traveling distributions. It is straightforward to modify equations (4) into traveling wave equations. Once again, we leave this subject to future exploration of the topic.

# V. MODEL LIMITATIONS AND PRACTICAL EXAMPLES

Throughout this paper we have neglected the spontaneous emission of radiation, or, in general, the quantization of the electromagnetic field. As was shown in some papers on Rabi oscillations (see, e.g., [7]), the semiclassical theory leads to almost the same qualitative predictions as the exact theory with field quantization, including the correct rates of spontaneous emission. The trapped mode solutions will be the same in the exact model in the limit of high stored energies in the trapped modes. In other words, the spontaneous radiation will be also trapped as well as the stimulated emission – there is no energy escape from the trapped modes even when the fields are quantized. When atomic motion is slow, and the ionization and other energy consuming effects are small, the fields can be trapped for very long times.

One more important physical property of the trapped modes should be discussed before we give estimations for available fields. Obviously, trapped fields exert pressure on the resonant atoms similar to the RF cavity case. To counteract the field pressure, we have to consider atomic restraining mechanisms in the medium. For example, the 1D atomic planes can be squeezed in between transparent solid slabs, or atoms can be placed in some crystal lattice in analogy with photonic crystals, forming some (not necessarily periodic) lattice. If the atoms are in a gas and the pressure from the fields is comparable with the gas pressure, in addition to given equation we must also solve the equation for pressure (or atomic density) that results from magnetic fields (since the magnetic field in our solutions is always in phase with currents). For steady state solutions with zero average velocities the equation reads:

$$\nabla P = f , \qquad (27)$$

where $P$ is the pressure and $f$ is the magnetic force, acting on the atomic currents. Obviously, in strong standing electromagnetic waves the atoms in the gas will form a diffraction grating for resonant atoms that will lead to even stronger trapping of waves. The total self consistent equation for the atomic density and the resonant fields will be treated in future explorations of this topic.

In principle, this electromagnetic field trap is similar to superconducting RF cavities. If we form (even temporarily) structures of standing or slowly moving atoms similar to RF cavities, the trapped modes can be used to provide acceleration. We now estimate the achievable fields in the trapped modes. The rotating wave approximation (RWA), used above, fails (see [8]) when the Rabi frequency becomes comparable with the transition frequency between two levels in our approximation, and, considering multi photon effects, the Rabi frequency should be less than the frequency difference between our chosen level and all closest levels. We take as an example the hydrogen atom and its levels with n=1 and n=2. The transition frequency is $\omega_0 \approx 1.55 \cdot 10^{16} Hz$ and $\mu_{12} = -\int d^3 r\, u_1^*(\vec{r})ez u_2(\vec{r}) = \frac{2^8 e a_0}{3^5 \sqrt{2}}$ for the transition between the 1st and 2nd states, where the upper level has *L*=1, and *m*=0 quantum numbers (the reference axis here is taken to be the direction of electric field), and $a_0$ is the Bohr radius. Assuming the Rabi frequency is only 1% of the transition frequency, the distance between levels *n=2 and n=3* is roughly fourteen times more than the Rabi frequency. Therefore, our approximation is reasonable for such values of the Rabi frequency. Equating the Rabi frequency $\Omega = \mu E_0 / \hbar$ to one hundredth of the transition frequency, we get a maximum amplitude (see the elaboration on maximum electric field after Equation (18)) of electric field $E_0 \sim 2.5$ *GeV/m*, that is almost 2 orders of magnitude larger than that of the achievable electric fields in superconducting cavities.

Unfortunately, the density of the Hydrogen gas must be very large. The trapped mode parameter $\chi = \dfrac{N k^2 \mu^2}{\varepsilon_0 \hbar \Delta}$ must be larger than $k^2$. After equating these and substituting the above $\mu = \mu_{12} = \dfrac{2^8 e a_0}{3^5 \sqrt{2}}$ into the expression for the parameter $\chi$, we obtain a minimum hydrogen density of $N \sim 3.3 \cdot 10^{21}$ cm$^{-3}$. This density is achievable at 100 atmospheres pressure. For some metals, the dipole transition coefficients can be an order of magnitude larger than that of hydrogen, therefore these fields can be achieved for pressures comparable with that of the atmosphere. For hydrogen at atmospheric pressure we obtain achievable fields of 25 MeV/m, which is also not small. But in any case, the total pressure of the fields can't be larger than the atmospheric pressure. For trapped modes with the radius $R=20$ cm the total energy of the state can be as large as $P \cdot 4\pi R^2 = 10$kJ.

We have neglected all energy consuming mechanisms, such as gas heating and gas ionization. However, for ultraviolet light, the air breakdown for wavelength of 0.1 μm requires a field of $10^9$-$10^{10}$ V/m ([9]). Therefore, the estimated fields can exist in the air without ionizing the gas. This model is one possibility to explain ball lightning – it may be a trapped electromagnetic wave whose frequency is close to the transition frequency between some atomic levels of the air atoms. In addition to trapping, some mechanism should exist to transfer energy from, for example, discharge in the air, to the trapped wave. These questions are beyond the scope of the present paper.

## VI. CONCLUSION

This paper demonstrates that there exist a variety of trapped solutions for electromagnetic fields in a medium where resonant atoms are present. The total reflection of electromagnetic waves by the medium is a necessary condition for their appearance. In addition to total reflection, the nonlinearity of atom-wave interaction is responsible for the phenomenon. Namely, for large intensities of the electromagnetic field, the "source term" (see RHS of Equations (15) and (20)) saturates and the medium becomes transparent to the wave. The region of "transparency" forms the core of the trapped mode. At the edges the field vanishes and the medium reflects the wave completely, thus confining the localized state with high electromagnetic energy density.

The trapped modes, in principle, may have zero loss and accumulate enormous field energy. It may be possible to use trapped modes in the same way as RF cavities or wave guides are used, but with larger achievable fields and with potentially zero energy loss. In addition, the trapped solutions for electromagnetic fields may be responsible for some atmospheric phenomena, such as "ball lightning".

## Acknowledgements

The author thanks A. Aleksandrov, J. Holmes, and A. Shishlo for useful discussions and help with manuscript proofreading. Research sponsored by UT-Batelle, LLC, under contract no. DE-AC05-00OR22725 for the U.S. Department of Energy.

## APPENDIX A. TRAPPED MODES IN RANDOM 1D LATTICE

Localized states also exist in random lattices. In addition to mode trapping, which is related to nonlinearity of field-atom interaction in the semiclassical approach, the randomness adds one more effect, known as "localization" in solid state physics [2]. We do not present a complete review the phenomenon here. However, we do point out one mathematical theorem by Furstenberg [5], which states that for any product of a large number of random noncommuting 1D (2×2) matrices with unit determinant, one of the eigenvalues of the resulting matrix will grow exponentially with the number of multiplication terms (the other eigenvalue is, of course, the inverse of the growing one). In application to our problem, a 1D random lattice of atoms will always yield exponentially growing (decaying) fields at infinity. Following Poincare (see, e.g. [4]), we define the sequence of points that decay exponentially at $-\infty$ as the $H^-$ curve, and the sequence of points, decaying at $+\infty$, as $H^+$, respectively. These curves are not trajectories, they are collections of trajectories leaving and returning to the center of phase space at infinities. In our case we have two $H^+$ and $H^-$ curves, each curve having its symmetric counterpart with respect to 180 degrees rotation of the phase space around the center. They are related to each other only by the fact that both have the same asymptotic growth/decay rate. The $H^-$ curve, in general, grows to substantial values (probably to infinity) at $+\infty$, as seen in numerical examples, but this growth is slow because the kick from each atomic node becomes constant for large values of the field (in our units, large as compared to unity). The $H^+$ curve behaves similarly, at $-\infty$. These curves, in general, are not the same for periodic systems, with the "small" exception of integrable systems. When the phase space, where these curves evolve, is confined or the growth is slow, as in our case, the curves cross each other with some nonzero angle (probably, there exist some exceptions, but we haven't seen them in our numerical cases). The crossing point is our trapped mode – it grows from zero values at $-\infty$ and decays exponentially at $+\infty$, because it belongs simultaneously to both "+" and "-"curves. The crossing of curves is confirmed by all numerical cases performed by the author (that, of course, does not constitute a mathematical proof). Here, we present one example of a random lattice with all the described curve and trajectory species, selected randomly, from infinitely many random cases.

For this case, we choose $\kappa=-0.8$, 100 random nodes and random phase advances from 0 to 1 radian (chosen as a fractional part of the node number after it is divided by $\pi$) between nodes. The approximate $H^-$ line was built in the following way: we selected a very small initial value of the forward field (see $u_f$ in equation (11)), evolved its growth through 10 nodes and connected first and last point by a line. Thus, we approximated the separatrix, coming emerging from zero for – ∞. We approximated this line by 1000 points placed uniformly over the line extent and tracked it over 50 nodes. The solid line in Figure 4 shows the resulting line in the "phase space" of the map, where the *x* coordinate stands for the imaginary part, and *y* – for the real part of the field. This line is an approximate piece of the H⁻ line. The same was done for the inverse map, starting from +∞, that corresponds to the 100$^{th}$ node, back to the center of the lattice (node 50). The dashed line in Figure 7 represents part of the H⁺ curve, obtained in this approximation. One can see that the curves do cross each other.

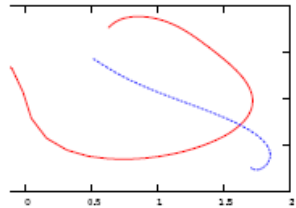

**Fig. 7.** Separatrix crossing.

The crossing point tracked from the 1$^{st}$ to the 100$^{th}$ node is shown as a trajectory in Figure 8 (dashed line). It starts at $-2.6*10^{-6}-i4.0*10^{-6}$ forward field value at node 100 and ends near zero at the 1$^{st}$ node (the actual values obtained are at $10^{-3}$ level, due to the limited accuracy of the numerical simulations and the strong influence of errors in the case with exponential instability). The solid line shows the trajectory of the H⁻ family that passes very close (within 0.01 at node 50) to the crossing point. Due to a slight error in initial conditions and the exponential instability, this trajectory eventually diverges from that of the trapped mode.

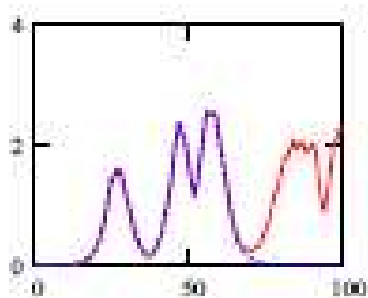

**FIG. 8.** Trapped mode shown by the dashed line. The horizontal coordinate corresponds to the node number, and the vertical coordinate shows the modulus of the forward wave. The solid line shows the field behavior for initial conditions 10% different from those of the dashed line.